# Resolving catastrophic error bursts from cosmic rays in large arrays of superconducting qubits

Matt McEwen,[1,2] Lara Faoro,[3] Kunal Arya,[2] Andrew Dunsworth,[2] Trent Huang,[2] Seon Kim,[2] Brian Burkett,[2] Austin Fowler,[2] Frank Arute,[2] Joseph C. Bardin,[2,4] Andreas Bengtsson,[2] Alexander Bilmes,[2] Bob B. Buckley,[2] Nicholas Bushnell,[2] Zijun Chen,[2] Roberto Collins,[2] Sean Demura,[2] Alan R. Derk,[2] Catherine Erickson,[2] Marissa Giustina,[2] Sean D. Harrington,[2] Sabrina Hong,[2] Evan Jeffrey,[2] Julian Kelly,[2] Paul V. Klimov,[2] Fedor Kostritsa,[2] Pavel Laptev,[2] Aditya Locharla,[2] Xiao Mi,[2] Kevin C. Miao,[2] Shirin Montazeri,[2] Josh Mutus,[2] Ofer Naaman,[2] Matthew Neeley,[2] Charles Neill,[2] Alex Opremcak,[2] Chris Quintana,[2] Nicholas Redd,[2] Pedram Roushan,[2] Daniel Sank,[2] Kevin J. Satzinger,[2] Vladimir Shvarts,[2] Theodore White,[2] Z. Jamie Yao,[2] Ping Yeh,[2] Juhwan Yoo,[2] Yu Chen,[2] Vadim Smelyanskiy,[2] John M. Martinis,[1] Hartmut Neven,[2] Anthony Megrant,[2] Lev Ioffe,[2] and Rami Barends[2]

[1]*Department of Physics, University of California, Santa Barbara CA, USA*
[2]*Google Quantum AI, Santa Barbara CA, USA*
[3]*Laboratoire de Physique Théorique et Hautes Énergies, Sorbonne Université, Paris, France*
[4]*Department of Electrical and Computer Engineering, University of Massachusetts Amherst, Amherst MA, USA*
(Dated: April 11, 2021)

Scalable quantum computing can become a reality with error correction, provided coherent qubits can be constructed in large arrays [1, 2]. The key premise is that physical errors can remain both small and sufficiently uncorrelated as devices scale, so that logical error rates can be exponentially suppressed. However, energetic impacts from cosmic rays and latent radioactivity violate both of these assumptions. An impinging particle ionizes the substrate, radiating high energy phonons that induce a burst of quasiparticles, destroying qubit coherence throughout the device. High-energy radiation has been identified as a source of error in pilot superconducting quantum devices [3–5], but lacking a measurement technique able to resolve a single event in detail, the effect on large scale algorithms and error correction in particular remains an open question. Elucidating the physics involved requires operating large numbers of qubits at the same rapid timescales as in error correction, exposing the event's evolution in time and spread in space. Here, we directly observe high-energy rays impacting a large-scale quantum processor. We introduce a rapid space and time-multiplexed measurement method and identify large bursts of quasiparticles that simultaneously and severely limit the energy coherence of all qubits, causing chip-wide failure. We track the events from their initial localised impact to high error rates across the chip. Our results provide direct insights into the scale and dynamics of these damaging error bursts in large-scale devices, and highlight the necessity of mitigation to enable quantum computing to scale.

Quantum states are inherently fragile. Superconducting qubits can achieve significant coherence only when cooled to milliKelvin temperatures and protected from their environment through extensive engineering. Maintaining coherence becomes an especially daunting task in quantum error correction (QEC), where all qubits must maintain low levels of physical error throughout lengthy algorithms. A correlated error event is particularly problematic, as affecting many qubits simultaneously is a recipe for logical faults [6].

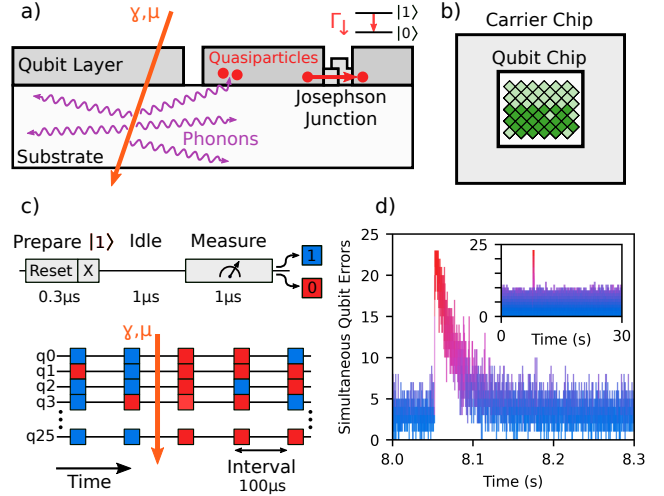

Figure 1. **Rapid repetitive correlated sampling.** (a) High energy radiation impinging on the device deposits energy, which spreads in the form of high energy phonons. In superconducting structures, this energy creates quasiparticles, which cause qubit energy decay as they tunnel across the Josephson junction. (b) We use a 26 qubit subset (dark green) of a Google Sycamore processor. The qubit chip is attached to a larger carrier chip using indium bumpbonds. (c) The Rapid Repetitive Correlated Sampling (RReCS) experiment consists of repeated cycles of preparation, idling and measurement. The idling time of 1 $\mu$s sets the sensitivity to decay errors. The interval between the start of each cycle is 100 $\mu$s. (d) A timeslice of a 30 second long dataset, showing an impact event. The number of simultaneous qubit decay errors jumps from baseline $\sim 4$ up to $\sim 24$, effectively saturating the chip. The number of errors returns to baseline with an exponential time constant of $\sim 25$ ms.

Energetic radiation is likely to produce such errors, as devices are bathed in a constant flux of high-energy rays [7]. Cosmic rays constantly impact the upper atmosphere, producing high-energy muons which can strike the chip directly or scatter in surrounding material to produce secondary rays. Additionally, gamma rays are commonly emitted from trace radioactive impurities, both directly and from Compton scattering of beta emissions. When interacting with the chip, these rays can deposit energies in the 100 keV to 1 MeV range [5, 8], dwarfing the typical energy scales in our qubits at around 25 $\mu$eV.

As illustrated in Fig. 1, incoming rays interact with matter along their path, ionizing the substrate and producing energetic phonons with long lifetimes. As they spread through the chip, these phonons can break paired electrons in superconducting structures and produce large densities of excess quasiparticles. A single quasiparticle may cause state decay as it tunnels across the junctions at the heart of superconducting qubits [9–13]. The avalanche of quasiparticles produced by such an impact has the ability to cause a chip-wide suppression of coherence [14].

High-energy radiation has been identified as a source of quasiparticles in transmon qubits [4], and single events have been reported in small numbers of specialized devices such as resonators [3] as well as qubits that are charge-sensitive [5]. However, observing these discrete events in detail presents a challenge in metrology. To understand the effect on QEC, one would need to observe how the events unfold in a device directly. This includes measuring the fast dynamics of the initial impact, the spread of errors through the qubit grid, and the eventual recovery to equilibrium. Therefore, a large array of qubits operated at rapid cycle times is required to illuminate the individual events and diagnose their impact on practical error correction.

Here, we directly measure the occurrence of high-energy events in a large-scale working device in the form of a Google Sycamore processor and provide insights into the microscopic dynamics of these events. We show that high-energy events produce discrete bursts of errors that affect an entire qubit patch on the processor, effectively lasting fo thousands of error correction cycles. Using fine time-resolved measurements, we show that events are initially localized but spread over the chip, providing strong evidence for a high-energy impact. Finally, we introduce a method to monitor the energy coherence time $T_1$ during an event and find it to be severely suppressed across all qubits, a clear signature of quasiparticle poisoning throughout the chip.

## RESULTS

### Rapid Repetitive Correlated Sampling

In order to measure these events in detail, we must rapidly identify correlated errors in large qubit arrays. We use a subset of a Google Sycamore Processor [15], as indicated in Fig. 1b. The qubit chip consists of an array of flux-tunable superconducting transmon qubits [16, 17] with tunable couplers [18–20]. Qubit operating frequencies are chosen algorithmically [21] between 6 and 7 GHz, with resulting $T_1$ at operating frequency around 15 $\mu$s. We turn off the coupling between neighbouring pairs of qubits. We operated only a subset of the device and some qubits with adverse calibration characteristics were not included, resulting in $N_Q = 26$ qubits being used. Each qubit lies around 1 mm from its nearest neighbours on a qubit chip measuring 10 mm x 10 mm, which is attached to a larger carrier chip measuring 20 mm x 24 mm using indium bumpbonds [22].

We introduce a method that rapidly and simultaneously measures qubit states to identify correlated errors, which we call Rapid Repetitive Correlated Sampling (RReCS). As indicated in Fig. 1c, all qubits are prepared in the $|1\rangle$ state, allowed to idle for a short sampling time (1$\mu$s), and then simultaneously measured. This cycle is repeated at rapid regular intervals (100$\mu$s) for extended periods of time, with any measurements where the qubit state has decayed to $|0\rangle$ recorded as an error. Finite $T_1$ and readout fidelities will produce errors that are independent between qubits, creating a low background error rate. With this technique, the quantum processor becomes a time-resolved detector for events that affect large numbers of qubits.

A time slice from a RReCS experiment is shown in Fig. 1d. It features a distinct peak where the total number of errors jumps from a baseline of $\sim$4 simultaneous errors up to $\sim$24 errors. This event has effectively saturated the qubit patch, with all qubits experiencing a high probability of reporting an error, indicating total failure of the coherence on the chip. The peak features an exponential decay back to the baseline error rate with a time constant around 25 ms, which is much larger than the typical QEC round time of 1 $\mu$s [23, 24]. The presence of such a long time period of elevated error rates would be unacceptable for any attempt at logical state preservation using QEC.

One signature of quasiparticle poisoning is an asymmetry between decay and excitation errors. Quasiparticles rapidly scatter and cool to energies near the superconducting gap $\Delta$, where they become unable to excite the qubit state from $|0\rangle \rightarrow |1\rangle$, which requires energy $\Delta + E_{01}$, where $E_{01}$ is the energy difference between the $|0\rangle$ and $|1\rangle$ states. However, quasiparticles maintain the ability to absorb the qubit energy and cause a decay error $|1\rangle \to |0\rangle$. As a test, we run the RReCS experiments for excitation errors, initialising $|0\rangle$ states and recording excitation to $|1\rangle$ states as an error. We do not find any correlated error peaks, indicating that events are produced by an asymmetric decay error mechanism, which is compatible with quasiparticle poisoning across the chip. Further detail on these experiments is included in the Supplementary Information.

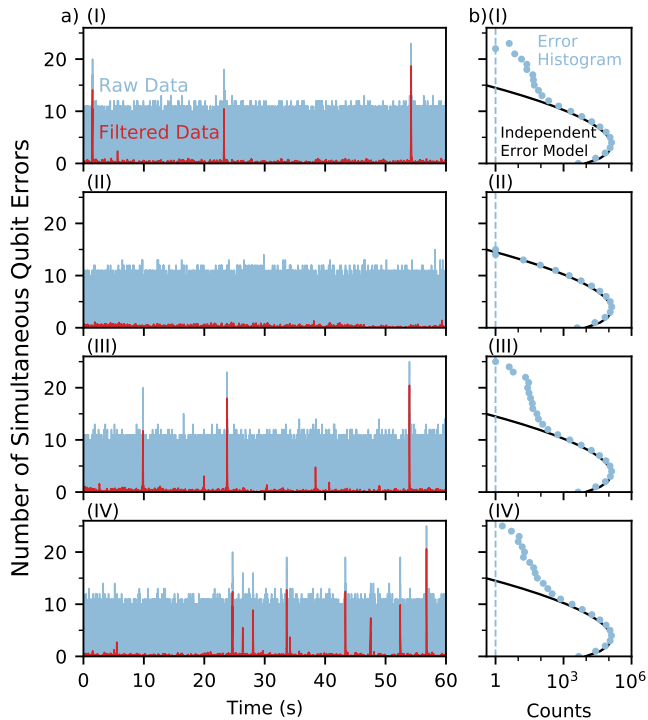

Figure 2. **Identifying events and background error.** (I-IV) Sequential datasets selected from a series of 100 datasets of 60 seconds each, with a time between data points of 100 $\mu$s. (a) Raw and filtered timeseries. Matched filtering allows for identifying events otherwise obscured by background noise. Events occur independently every 10 seconds on average. (b) Histograms of the number of errors, along with a model of independent errors arising from qubit $T_1$ and readout. In the absence of events the data closely correspond to this model, whereas when events occur high error counts appear.

## Timing of Events and Independent Background Error

To understand the arrival rate and uniformity of impact events, we now deploy RReCS experiments for long time periods to gather large numbers of events. We acquire 100 back-to-back datasets of 60 seconds each, and apply a matched filter to isolate events over the background independent error rate. Details on this filtering are included in the Supplementary Information. Four sequential datasets are shown in Fig. 2, selected to include one dataset without any events present. In Fig. 2a, the raw time series data illustrates the background error rate, but the filtered data displays low noise and clearly identifies events even at scales lower than the background noise level. Fig. 2b shows corresponding histograms over the number of simultaneous errors, where the black lines indicate the expected background distribution of independent errors.

We include a simple independent error model, where we assume perfect initialization, followed by population decay with an independently measured $T_1$ time over the 1 $\mu$s sampling time, and finally account for separately measured finite readout fidelities. In the absence of events, we note a strong correspondence of the background error distribution to this simple model, as illustrated in Fig. 2b (II). In the presence of events, we note a distinct excess of high numbers of simultaneous errors, well above what is reasonable for uncorrelated error sources. This indicates that the baseline performance of the experiment is well understood and that the peaks represent anomalous correlated error events.

Using our matched filter, we extract 415 events from these datasets, which we then fit individually to extract a peak height and exponential decay timescale. Details on this analysis and the distributions of extracted parameters are included in the Supplementary Information. We find that the decay timescale is tightly grouped in the 25 to 30 ms range, and that peak heights range from the minimum identifiable by our analysis up to the full number of qubit used. We also extracted 326 time periods between events occurring in the same dataset and find a strong correspondence to an exponential distribution with an average event rate $\lambda = 1/(10 \text{ s})$. This indicates that the occurrence of the events is independent over time, occurring on average every 10 s without significant bunching or anti-bunching. This timescale is long compared to the typical qubit coherence times, and therefore will have a limited influence on typical qubit $T_1$ measurements [4]. However, this timescale is quite short compared to the run time of error corrected algorithms, which is projected to be several hours [25], so any attempt to preserve a logical state for computation is very likely to see such an event.

## Impact Localization and Evolution

We now turn to experiments with higher time resolutions in order to observe the evolution of individual events as they progress. Our use of a recently developed reset protocol [23] was key in allowing us to achieve 3 $\mu$s intervals between measurements and thereby acquire resolution inside the rising edge of the event. In Fig. 3a, saturation regime at 1.5 ms following the impact, where the hot spot has grown in size and all of the qubit patch sees noticeably elevated error rates. Finally, (4) shows the performance 10 ms after the event during the exponential tail. The initial impact site is still visible but less distinct from the surrounding area. Error rates throughout the chip are still noticeably elevated above baseline levels. Two further events at this level of time resolution are included in the Supplementary Information for comparison, each featuring the same timescales and showing clear localization around different points on the chip.

This is direct evidence of a localised impact on the device which induces high error rates as it spreads over the chip. It further identifies the relevant timescales for the dynamics associated with the initial impact and spread, providing the first insights into the device physics underlying these processes. The spread of the initial hot spot arises from the interplay between phonons traveling through the substrate before being absorbed, and the rate of quasiparticles recombining and emitting further phonons in the qubit layer. At the start of this process, we estimate a timescale of $\sim 180$ $\mu$s. As the event continues, the energy will be distributed over a larger volume, slowing the rate of recombination, and explaining the qualitative shape of the rising behaviour seen here. We include details on these interactions and estimates in the Supplementary Information. These timescales will be also influenced by additional structures and materials on the chip, including the indium bumpbonds and second substrate in a flip chip device. As devices continue to grow in size and complexity, and choices of materials become more diverse, one could expect to see more non-trivial dynamics in response to high-energy stimuli.

### Quasiparticle Signature and Event Magnitude

One key signature of a quasiparticle poisoning event is a suppressed qubit $T_1$, producing an exponential relationship between error rate and sampling time. For a chip-wide event, we should also see that the absolute number of errors is predicted by a limited $T_1$ across all qubits. Here, we use the speed of our approach to perform Time-multiplexed RReCS (T-RReCS), rapidly cycling through sequential measurements. Each measurement probes for a varying sampling time, between 2 $\mu$s and 0 $\mu$s as shown in Fig. 4a. We separate this data into 5 concurrent timeseries, one for each sampling time.

We show a time slice from one such experiment in Fig. 4b. To extract an effective $T_1$ during the event, we first perform an exponential fit to find the height of the peak for each sampling time. While this fit neglects the rising behaviour shown previously, it is dominated by the long 25 ms tail of the event and produces an accurate estimate of the number of errors at the peak of the event. Fig. 4c shows the fitted heights plotted versus we show the raw time trace focusing on the start of an event, with Fig. 3b showing the longer tail of the event. We find three distinct timescales; an immediate jump in error from baseline at $\sim$4 errors to $\sim$10 errors in only $\sim$10 $\mu$s, a slower saturation up to a maximum of $\sim$15 over the following $\sim$1 ms, and a typical $\sim$25 ms exponential decay back towards baseline. Fig. 3c shows heatmaps of the errors over the device averaged over a 300 $\mu$s window: (1) shows the baseline performance starting 400 $\mu$s prior to the impact, displaying homogeneous low error rates. (2) includes the immediate jump to elevated error rates, displaying a localized hot spot where the highest error rates are concentrated. (3) shows the end of the sampling time. We fit the heights to a two-parameter exponential model $h = a(1 - \exp(-t_{\text{sampling}}/T_1^{\text{avg}}))$, where $h$ is the height of the fitted peak, and $t_{\text{sampling}}$ is the sampling time. The timescale $T_1^{\text{avg}}$ extracted from the relative peak heights is analogous to an average $T_1$ over the device. We then compare this extracted $T_1^{\text{avg}}$ to a simple global $T_1$ model for predicting the absolute peak heights, as shown in Fig. 4d. Each vertical group of points represents 5 heights extracted from a single event. The absolute peak heights correspond well to a zero-parameter model; $h = N_Q(1 - \exp(-t'_{\text{sampling}}/T_1^{\text{avg}}))$, corresponding to the expected number of qubit errors for $N_Q = 26$ qubits with identical $T_1$ equal to the $T_1^{\text{avg}}$ extracted from the relative peak heights. Here, $t'_{\text{sampling}} = t_{\text{sampling}} + 500$ ns accounts for the finite readout time, as any decay during the first half of measurement integration of 1 $\mu$s will be recorded as an error.

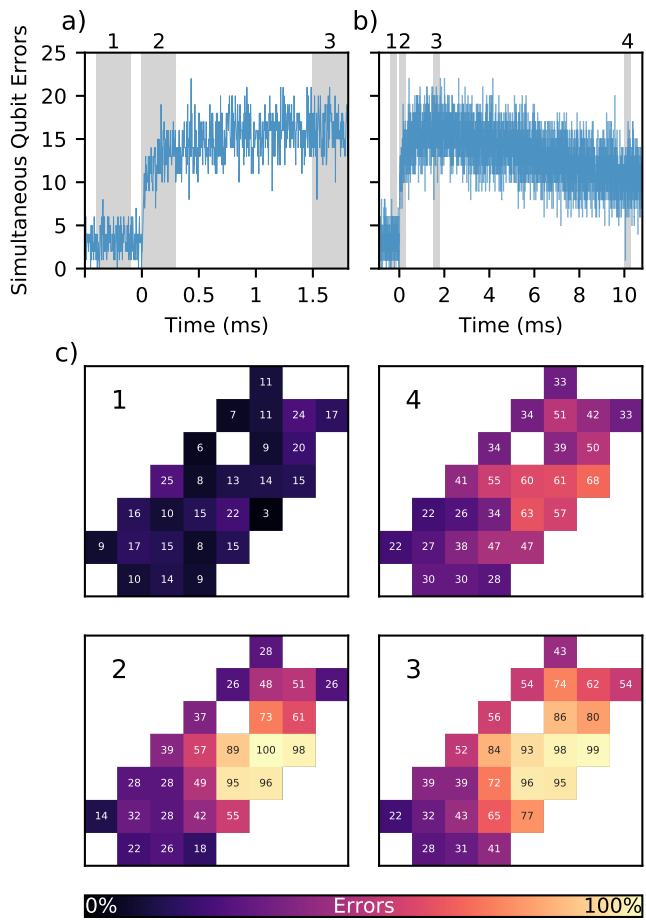

Figure 3. **Localization and spread of error.** (a-b) Time-slices of the same event taken from a dataset with a sampling time of 1 $\mu$s and an interval of 3 $\mu$s between datapoints. The number of errors jumps up from a baseline of $\sim 4$ to $\sim 10$ errors in around 10 $\mu$s, then rises to $\sim 15$ errors in around 1 ms, before returning to the baseline following an exponential decay with a time constant of $\sim 25$ ms. (c) Heatmaps of the qubit patch, showing the error rate in percent averaged over 300 $\mu$s slices located (1) before the event, (2) at the initial impact, (3) after the rise to the peak value, (4) during the recovery of equilibrium. High error rates are initially localized to a small number of qubits, but spread through the device over the course of the event.

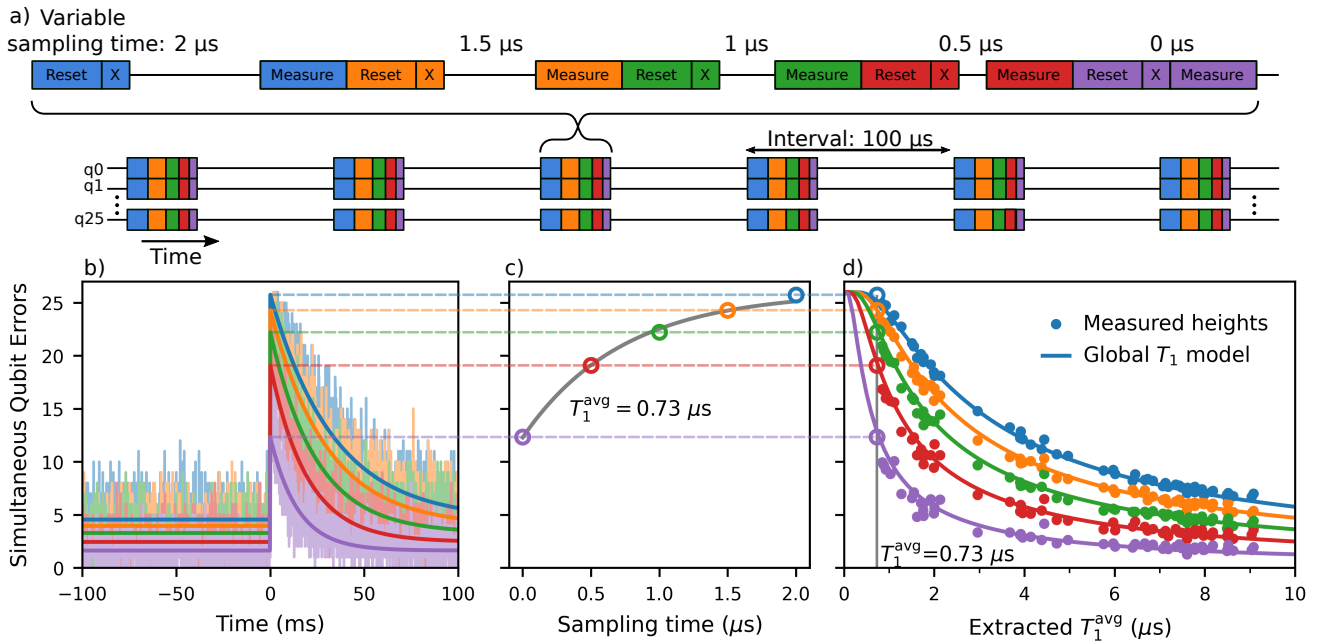

Figure 4. **Extracting energy decay times during events.** (a) A time-multiplexed rapid repetitive correlated sampling (T-RReCS) experiment, where each cycle consists of five measurements of varying sampling time. (b) A single event shown in five concurrent timeseries with different sampling times, with truncated exponential fits to extract peak heights. (c) Peak heights $h$ vs. sampling time $t_{\text{sampling}}$, showing an exponential fit to $h = a(1 - \exp(-t_{\text{sampling}}/T_1^{\text{avg}}))$, showing that the peak heights scale exponentially with sampling time, as expected for quasiparticle-limited $T_1$ times. (d) Peak heights $h$ plotted vs extracted $T_1^{\text{avg}}$, along with a global $T_1$ model having no free parameters, $h = N_Q(1 - \exp(-t'_{\text{sampling}}/T_1^{\text{avg}}))$. We take $N_Q = 26$, equal to the number of qubits in the experiment, a homogeneous $T_1 = T_1^{\text{avg}}$, and use $t'_{\text{sampling}} = t_{\text{sampling}} + 500$ ns to account for the finite measurement time of 1 $\mu$s. This shows the events can be summarized by a chip-wide drop in $T_1$ that can be inferred directly from the peak heights. Grey line indicates the decay time extracted in (c).

This shows that events can be summarized by a global drop in effective $T_1$ during an event, providing direct evidence of decay errors induced by excess quasiparticles in the qubits, rather than by coherent control or readout failure. Further, the magnitude of the $T_1$ drop can be inferred directly from the peak height. While there is localization on short timescales, the events generically affect all qubits enough to correspond well to a simple homogeneous model. The largest detected events correspond to a chip-wide reduction in $T_1$ to less than 1 $\mu$s.

## DISCUSSION

We have shown direct measurements of widespread, long-lived and highly detrimental correlated events in a practical quantum processor, presenting an existential challenge to quantum error correction. Our results provide a natural explanation for the observations of short periods of significantly elevated error found in recent experiments with stabilizer codes [23, 24]. Our findings are also relevant for other solid-state quantum systems, such as spin qubits that would be sensitive to the induced charges [26], and Majorana fermion devices that would be sensitive to induced quasiparticle population [27, 28].

While the incidence rate can be reduced by shielding [3, 4], the ability of even a single event to cause widespread failure calls for mitigation efforts on the chip itself [14]. We are encouraged by recent works on quasiparticle and phonon trapping in qubits [29–31], and mitigation efforts in astronomical detectors using membranes and phonon down-converters [8]. Adopting these approaches successfully for error correction will rely on an understanding of the detailed evolution of the events, especially as devices increase in size and complexity. We hope that our work spurs research into the physics of correlated events and accelerates the development of mitigation efforts that help enable quantum error correction at scale.

## AUTHOR CONTRIBUTIONS

M.M. and R.B. designed the experiments. M.M. performed the experiments and analysed the data. L.F., L.I. provided theoretical models. K.A., M.M. implemented key components of the data taking infrastructure. A.D., T.H., S.K., B.B. designed and fabricated relevant devices. M.M., L.F., L.I., R.B. prepared the manuscript. All authors contributed to the experimental infrastructure and manuscript revision.

## ACKNOWLEDGEMENTS

The authors would like to acknowledge stimulating discussions with B. Mazin, J. Baselmans, A. Endo, K. Karatsu, R. McDermott and C. Wilen.

## COMPETING INTERESTS

The authors declare no competing interests.

## DATA AVAILABILITY

The data that support the findings of this study are available from the corresponding author upon reasonable request.

# Supplementary information for "Resolving catastrophic error bursts from cosmic rays in large arrays of superconducting qubits"

Matt McEwen,[1,2] Lara Faoro,[3] Kunal Arya,[2] Andrew Dunsworth,[2] Trent Huang,[2] Seon Kim,[2] Brian Burkett,[2] Austin Fowler,[2] Frank Arute,[2] Joseph C. Bardin,[2,4] Andreas Bengtsson,[2] Alexander Bilmes,[2] Bob B. Buckley,[2] Nicholas Bushnell,[2] Zijun Chen,[2] Roberto Collins,[2] Sean Demura,[2] Alan R. Derk,[2] Catherine Erickson,[2] Marissa Giustina,[2] Sean D. Harrington,[2] Sabrina Hong,[2] Evan Jeffrey,[2] Julian Kelly,[2] Paul V. Klimov,[2] Fedor Kostritsa,[2] Pavel Laptev,[2] Aditya Locharla,[2] Xiao Mi,[2] Kevin C. Miao,[2] Shirin Montazeri,[2] Josh Mutus,[2] Ofer Naaman,[2] Matthew Neeley,[2] Charles Neill,[2] Alex Opremcak,[2] Chris Quintana,[2] Nicholas Redd,[2] Pedram Roushan,[2] Daniel Sank,[2] Kevin J. Satzinger,[2] Vladimir Shvarts,[2] Theodore White,[2] Z. Jamie Yao,[2] Ping Yeh,[2] Juhwan Yoo,[2] Yu Chen,[2] Vadim Smelyanskiy,[2] John M. Martinis,[1] Hartmut Neven,[2] Anthony Megrant,[2] Lev Ioffe,[2] and Rami Barends[2]

[1]*Department of Physics, University of California, Santa Barbara CA, USA*
[2]*Google Quantum AI, Santa Barbara CA, USA*
[3]*Laboratoire de Physique Théorique et Hautes Énergies, Sorbonne Université, Paris, France*
[4]*Department of Electrical and Computer Engineering, University of Massachusetts Amherst, Amherst MA, USA*
(Dated: April 11, 2021)

## I. ENERGY CASCADE FOLLOWING A HIGH ENERGY IMPACT

A high-energy impact produces an avalanche of excitations as the energy spreads through the system. Impacts are severe, depositing energy at much larger scales than those characteristic of our qubits and chip materials. As this energy quickly cascades from the initially small number of high energy particles to lower energies, it produces an explosion of excitations in the device.

This cascade progresses through several distinct stages as the event evolves [1]. The primary interactions in this avalanche are the initial deposition of energy by the incident radiation and the recombination of resulting charges, the propagation and down-conversion of phonons, and finally the creation and eventual recombination of quasiparticles in the qubits. We will discuss these in turn.

The impact event starts with the high energy particle crossing the chip and depositing a large amount of energy. The two particles of most concern are gamma rays from radioactivity in materials surrounding the chip, and muons produced by cosmic ray showers. Gamma rays are more common, depositing energy in the substrate via Compton scattering. Muons are more rare but deposit higher energies, leaving an ionization track along their path through the device. For a smaller sample with area 40 mm$^2$ gamma ray and cosmic ray events where reported with a rate of $\sim 1/(100\,\text{s})$ and $\sim 1/(500\,\text{s})$, and deposited energies around $\sim 100$ keV and $\sim 1$ MeV respectively [2]. For our carrier chip of 20 mm x 26 mm, the equivalent event rates are $\sim 1/(7.6\,\text{s})$ and $\sim 1/(38\,\text{s})$ respectively.

These impacts create electron-hole pairs in the insulating substrate of the device. The typical energy for an electron-hole pair is $E_{\text{e-h}} = 3.75$ eV [3], so the number of charges created by an event is $\sim 10^5$. The majority of electron-hole pairs immediately recombine and emit photons but $\sim 20\%$ of them avoid immediate recombination and propagate in the substrate [2]. These remaining free charges create further high energy phonons as they slow down to the speed of sound. In Si, the mobility of electrons is higher than that of holes, so typical events generate a net current of electrons resulting in the $1/f^2$ charge noise observed in charge sensitive transmons [4]. Slow charges eventually get trapped by ions and other defects in the substrate, with a characteristic length of $\lambda_{trap} \sim 300\,\mu$m in Si [5]. Charges that arrive at superconducting materials are trapped at much shorter distances. They quickly emit phonons as they decrease in energy down to the superconducting gap and may get trapped in local variations of the gap. Such trapped charges act as effective two-level systems, and can be characterized by a non-uniform density of states [6].

Phonons produced by the event are primarily responsible for the spread of energy through the chip. High-energy phonons downconvert to lower energies as they propagate through the substrate, through the phonon-phonon interaction arising from the anharmonicity of the substrate. Initially, the high-energy phonons quickly downconvert to acoustic phonons with energies below the Debye energy ($\hbar\omega_D = 56$ meV for Si). This phonon decay rate however slows down at lower energies with

$$\gamma_{\text{insulator}}(E) \approx g\frac{E}{\hbar}\left(\frac{E}{\hbar\omega_D}\right)^4, \quad (S1)$$

where $g \approx 0.01$ is the phonon anharmonicity for a typical insulator [7]. During the early stages of the impact ($\sim 100\,\mu$s), phonons propagating through the substrate can decay down to energies near $\sim 5$ meV, still significantly higher than the superconducting gap of aluminum $2\Delta \approx 0.36$ meV.

The absorption of phonons by metallic structures is the primary mechanism for further down conversion. In normal metals, acoustic phonons produce electron-hole pairs. Each particle in the pair emits low energy phonons, resulting in an efficient energy downconversion process with decay rate $\gamma_{\text{metal}}(E) \propto E$. In superconductors, the phonon absorption is similar to that in normal metals above the superconducting gap $E \gtrsim 2\Delta$ but ceases below it. The resulting quasiparticles similarly cool rapidly to near the superconducting gap by emitting phonons.

The spatial extent of the initial hotspot is also determined by the absorption of phonons by superconductors on the chip. The chip features an aluminum groundplane and qubit layer 100 nm thick on a silicon substrate around 500 $\mu$m thick, and also features significant volumes of indium in the form of bump bonds. These bump bonds are around 5 $\mu$m tall and cover $\sim 15\%$ of the surface area of the qubit chip. Indium features a superconducting gap $\Delta_{\text{In}} \approx 0.52$ meV, nearly three times higher than aluminum.

The rate at which a phonon traveling in a superconductor will break pairs and produce quasiparticles is given by [1, 8]

$$\tau_b^{-1} \approx \frac{1}{\tau_0^{ph}}; \qquad E_{ph} \sim 2\Delta \qquad (S2)$$

$$\tau_b^{-1} \approx \frac{1}{\pi\tau_0^{ph}}\frac{E_{ph}}{\Delta}; \qquad E_{ph} \gg 2\Delta \qquad (S3)$$

where $\tau_0^{ph}$ is the characteristic phonon lifetime, which is proportional to $1/\Delta$ and for aluminum is $\tau_0^{ph} \approx 0.24$ ns [8]. First, we discuss the aluminium layer alone. Considering phonons downconverted by the substrate to $E_{ph} = 5$ meV, we find $\tau_b \approx 27$ ps. The absorption length is then $l_{ph} = c\tau_b \approx 170$ nm, where $c = 6.4$ km/s is the speed of sound in aluminum. This distance is comparable to the aluminum thickness, so most of the high energy phonons radiating from the initial impact will be absorbed locally by the aluminum. The initial quaiparticles will cool rapidly toward the superconducting gap, radiating further phonons with lower energies. These phonons will ahve energies ranging from the initial energy of the quasiparticle to near the superconducting gap. At a moderately lower energy of $E_{ph} = 1$ meV close to the superconducting gap of indium, we find a longer pairbreaking time $\tau_b \approx 130$ ps and absorption length $l_{ph} \approx 870$ nm. At the aluminum gap $E_{ph} = 2\Delta \approx 0.4$ meV, we find $\tau_b \approx 240$ ps and $l_{ph} \approx 1500$ nm. The reduced absorption at these lower energies would result in significant spread of the initial hotspot when considering the aluminum layer alone.

However, the large volumes of indium will also absorb phonons with energies above the indium superconducting gap. Even at the lowest relevant phonon energy $E_{ph} = 2\Delta_{\text{In}} \approx 1$ meV, the phonon lifetime in indium is only $\tau_0^{ph} \approx 170$ ps. Given the speed of sound is $c \approx 1.2$ km/s, the phonon absorption length near the gap is $l = c\tau_0^{ph} \approx 200$ nm. Given the thickness of the bump bonds, any phonons above $E_{ph} = \Delta_{\text{In}}$ impinging on the indium bumps will be absorbed. This additional absorptivity will contribute to tighter localization of the initial hotspot and will slow the effective spread of phonons until they are below $2\Delta_{\text{In}}$. However, as the aluminum will downconvert phonons below the indium superconducting gap, we expect the rate of spread to increase as the event progresses.

The spread of error through the chip is also influenced by the rate of recombination in the superconductors. The recombination rate of quasiparticles that have cooled to near the gap is proportional to the quasiparticle density [8],

$$\tau_r^{-1} = \frac{2\pi\Delta\alpha^2(2\Delta)F(2\Delta)}{Z\hbar}x_{qp} \qquad (S4)$$

with $\alpha^2(2\Delta)F(2\Delta)$ the Eliashberg function at the gap energy, $Z$ a renormalization parameter, and $x_{qp} = n_{qp}/n_{cp}$ the normalized density of quasiparticles. The above can be approximated as $\tau_r^{-1} \approx x_{qp} \cdot 21.8/\tau_0$ [1], with $\tau_0 = 440$ ns the characteristic quasiparticle time for aluminum [8]. Again, we will first consider the aluminum layer only. For a median impact delivering 100 keV of energy to the substrate [2], where 20% of the energy remains in free charges, the energy absorbed by the initial hotspot in the aluminum layer will be around $E_{hs} = 80$ keV. We take area of the hotspot to be $A = 10$ mm$^2$, similar to that seen in segment 1 of Fig. 4 of the main paper. The aluminum thickness $d = 100$ nm. For aluminum, $\Delta \approx 0.18$ meV and $n_{cp} \approx 4 \times 10^6$ $\mu$m$^3$. We calculate the density of quasiparticles $x_{qp} = E_{hs}/Ad\Delta n_{cp} \approx 1.1 \times 10^{-4}$. The recombination time is then $\tau_r \approx 180$ $\mu$s, which agrees well with the experimentally observed time for the expansion of the initial hotspot. As the energy spreads through the device, the resulting quasiparticle densities will be lower and this recombination time will grow.

We now consider the recombinaiton in the indium. Indium features a very high diffusivity of quasiparticles, meaning any induced quasiparticle density will distribute itself throughout the bumpbond volume quickly. The bumpbonds are fabricated on top of the aluminum groundplane, separated by a thin titanium nitride (TiN) diffusion barrier. The superconducting gap of TiN films depends sensitively on the thickness and presence of contaminants, and there is evidence in literature for the possibility of a smeared gap that might allow conduction [9].

In the case that the TiN film has a lower superconducting gap than the indium, quasiparticles will rapidly flow from the indium into the aluminum, where they will cool to the aluminum gap. Unable to flow back into the indium, they will recombine in the aluminum at around the rate found previously. If the TiN film has a higher gap and features no subgap conduction, then we can similarly estimate the recombination time for the quasiparticles trapped in the indium. The initially absorbed phonons will distribute energy between the aluminum and indium by their exposed surface area and rate of absorbtion. Taking the same hotspot area of 10 mm$^2$, and considering an energy absorbed in the indium to be $E_{\text{In}} = 50$ keV, the recombination timescales are similar. In indium, $n_{cp} \approx 13 \times 10^6$ $\mu$m$^3$, so we find a small quasiparticle density $x_{qp} \approx 1 \times 10^{-6}$ due to the large volume. The short characteristic phonon timescale $\tau_0^{ph} = 0.799$ ns results in a recombination time $\tau_r \approx 36$ $\mu$s. However, only recombinations taking place near the surface of the indium will be able to radiate out. Accounting for ratio of the phonon absorption length $l \approx 200$ nm and the height of the bumpbonds $z \approx 5$ $\mu$m, we calculate the effective relaxation time of the indium bumps as $\tau_r^{\text{eff}} = \tau_r z/l \approx 0.9$ ms. This timescale is similar to the spread of energy from recombination in the groundplane, so recombination in indium is also compatible with the observed spread of the hotspot. In the case where a lower proportion of energy is trapped in the indium, the recombination time scale grows larger than the contribution from aluminum and will contribute to the leading dynamics of the spread. As such, the indium may play a role in the initial absorption of energy. However, the presence of the aluminum will soon downconvert phonons to below the indium gap, effectively freezing the indium out of the dynamics at longer timescales.

The presence of quasiparticles near the junctions in the qubits are the cause of the observed errors. The relaxation rate of the qubits is proportional to the density of quasiparticles [10]:

$$\frac{\Gamma_1}{\omega_q} = \sqrt{\frac{2\Delta}{\pi^2 \hbar \omega_q}} x_{qp} \tag{S5}$$

In Fig. 4 of the main paper we extracted an effectively global $T_1$ time around 1 $\mu$s at the peak of large events, which is compatible with estimates based on similar device structures [1]. As a consistency check, we can extract the total energy necessary in the qubit layer to induce this $T_1$ time. Using a typical qubit frequency $\omega_q = 2\pi \times 6$ GHz, we calculate a quasiparticle density of $x_{qp} \approx 2.1 \times 10^{-5}$ at the peak of the event, once the error has spread over a large area. We can convert this to an absolute density of $n_{qp} = 87$ $\mu$m$^{-3}$. Producing this density evenly over the full aluminum qubit layer on the 10 mm x 10 mm qubit chip would require an absorbed energy in the aluminum of $E = n_{qp} V \Delta \approx 160$ keV, which corresponds well to the expected energy depositions from large events.

Finally, the energy must eventually escape the chip through thermalization with the surrounding cryostat. The chip is floating in the package, suspended by aluminum wirebonds. The observed $\sim 25$ ms timescale for the return of the device to equilibrium is in reasonable agreement with a simple model of thermalization that gives $\sim 8$ ms [1]. Given the shorter timescale for this relaxation found in Ref. [2], we expect that increased thermalization will serve to reduce the overall duration of the event.

These interactions can help us to understand the error mechanism observed in this work, but we neglect some of the intricacies involved in the energy cascade in a practical device. Impacts depositing different initial energies will influence the details of the induced densities and timescales of recombination. Impacts in different locations across the device will also influence the exposure of the qubit patch to the initial hotspot and to the spreading of errors. Finally, the interplay between other structures and materials on the chip will also introduce additional complexity, including the distribution of energy between the indium and aluminum, the behaviour of titanium nitride diffusion barrier and the presence of a second substrate as carrier in flip-chip devices. A quantitative understanding of the energy cascade in such large devices remains an open question, and will be important for guiding efforts to mitigate events by introducing structures that affect the energy cascade.

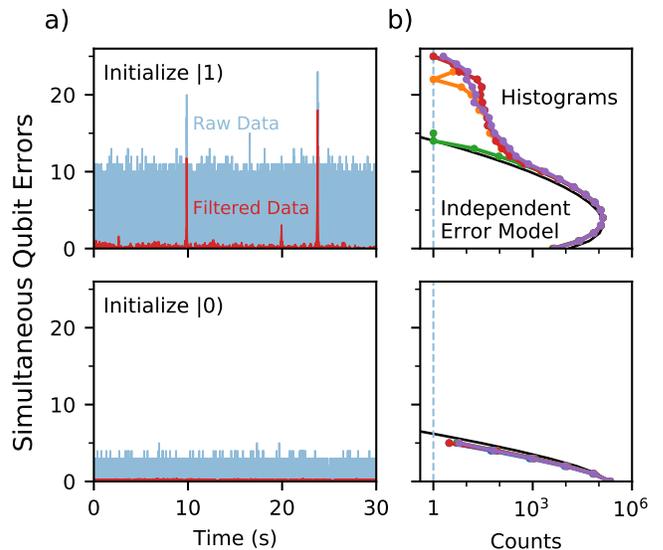

Figure S1. **Decay vs Excitation Errors** (a) Timetraces for RReCS experiments which initialise $|1\rangle$ and $|0\rangle$ respectively. Baseline error rates are much lower when initialising $|0\rangle$. (b) Histograms for 5 datasets (different colors) initialising $|1\rangle$ and $|0\rangle$ respectively. No datasets initializing $|0\rangle$ display any high-number counts that indicate a correlated error event.

## II. ABSENCE OF EXCITATION ERRORS

A key characteristic of quasiparticle poisoning in our qubits is the predicted asymmetry between excitation and decay errors. To measure this asymmetry, we also perform RReCS experiments looking for excitation errors by preparing qubits in $|0\rangle$ and recording measurements of $|1\rangle$ as errors. Fig. S1a shows timetraces for both a regular RReCS experiment initializing $|1\rangle$ and for excitation-sensitive RReCS experiment initializing $|0\rangle$, including both the raw timeseries and matched filtered timeseries. We note that initializing $|0\rangle$ produces much lower background error rates, as $T_1$ error is suppressed. We find a lack of correlated events, especially clear in the low noise matched filtered timeseries. Fig. S1 shows histograms for 5 such datasets of each type. We see that the histograms when initializing $|0\rangle$ correspond closely to the independent error model, and do not feature the elevated counts of high-error-number points characteristic of the events found when initializing $|1\rangle$. Over more than 35 of such datasets, we found no evidence of any correlated events in excitation.

This represents further evidence that the error mechanism responsible for the correlated error events is high densities of quasiparticles. Quasiparticles in the superconductor rapidly cool to energies near the superconducting gap $\Delta$. When tunneling across the Josephson junction, these quasiparticles are able to absorb the qubit energy $E_{10}$, producing a decay error. Once cooled to near the gap, quasiparticles are not capable of exciting the qubit state, as this requires energies of at least $\Delta + E_{10}$. As such, excitation errors are not expected to be present in chip-wide quasiparticle poisoning, as they would be in the case of an induced chip-wide readout or coherent control failure. The strong asymmetry in the rates of correlated decay and excitation errors is therefore strong evidence that the errors in events are due to excess quasiparticles.

## III. PEAK-FINDING AND EVENT PARAMETER DISTRIBUTIONS

Given the large volume of data that long RReCS experiments produce, it is important to be able to identify events in a reliable and automated fashion. Fig. S2a shows time slice of raw data including an event, illustrating the sharp jump and exponential decay that is characteristic of all events. Because the event shape is reliable, we can use a matched filter to optimally reduce noise and locate peaks. Throughout, we will use a template function:

$$\text{Template}(t) = \begin{cases} a \exp(-(t-t_0)/\tau_{\text{decay}}) + c & ; t \geq t_0 \\ c & ; t < t_0 \end{cases}$$

We remove the DC component of the time series, and then apply a matched filter created from the template with the values $\tau_{\text{decay}} = 20$ ms, $a = 1$, $c = 0$, $t_0 = 0$. Fig. S2b shows the resulting filtered function, illustrating the low noise level and symmetric peak produced by the matched filter. Our selection of $\tau_{\text{decay}} = 20$ ms will influence the symmetry and scale of the resulting peak, but not the location of the maximum. We then apply a threshold of 2 errors to the filtered time series and return local maximum for each section over the threshold to identify peaks. This technique is capable of detecting peaks of heights greater than 2 errors over the background error value around 4 errors, with the smallest identified peaks having absolute heights of 6 errors. This technique could be easily refined to capture events with heights even closer to baseline due to the reliability of the shape of the event even at very small scales.

To extract parameters for each event, we return to the raw data to make a direct fit. We perform a least squares fit using the template given above to the time-series around the located peak. Fig. S2a shows the resulting fit over the raw time series, from which we extract the peak time $t_0$, the decay constant $\tau_{\text{decay}}$ and the peak height $h = a + c$. Fig. S2c shows the distribution of fitted decay constants $\tau_{\text{decay}}$ for the 415 peaks found in the 100 datasets presented in the main paper, showing a tight distribution around 25-30 ms. Fig. S2d shows the distribution of heights $h = a + c$ for the same peaks. We see events from the minimum size identifiable of 6 errors at peak, up to the full size of our qubit patch at $N_Q = 26$ simultaneous errors. Smaller events are found more frequently, qualitatively matching the distribution of deposited energies for impinging particles [2].

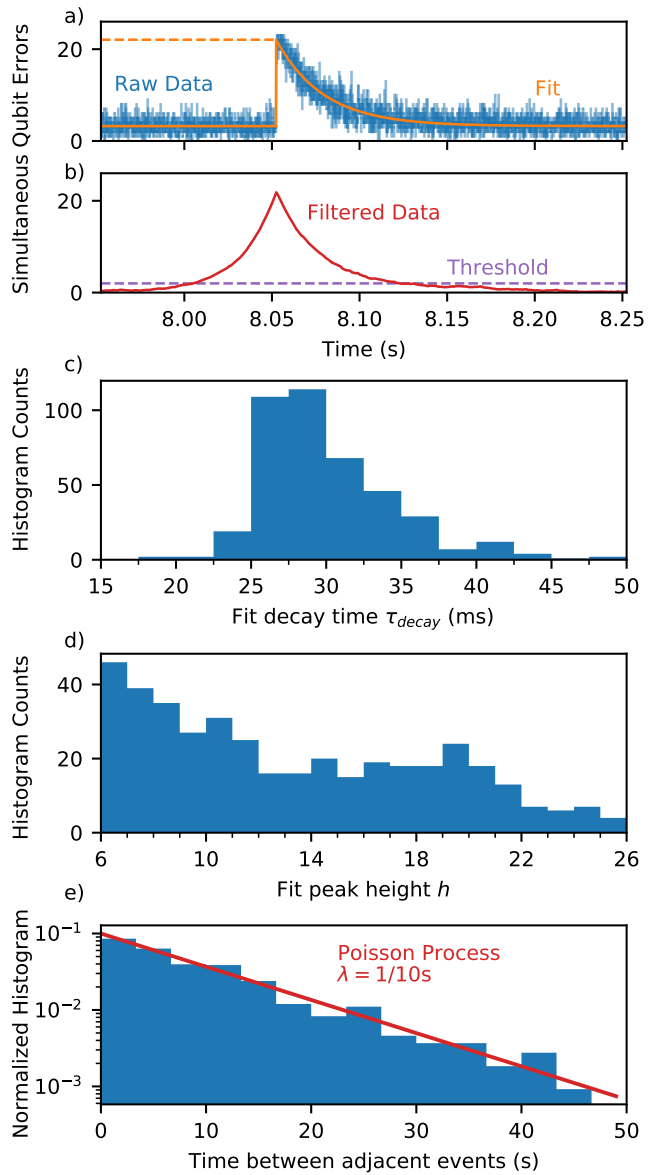

Figure S2. **Peak-finding procedure and parameter distributions** (a) A time slice including a single event, showing the raw data (blue) and the final least squares fit used to determine peak parameters (orange). (b) The same time slice after application of a matched filter (red), including the threshold at 2 errors above mean (purple) used to locate peaks for fitting. (c) The distribution of exponential decay times $\tau_{\text{decay}}$ fit to events, showing a tight grouping around 25-30 ms. (d) The distribution of fit peak heights. The minimum peak height possible to identify with this analysis method is around 6, and the maximum is the full number of qubits at 26. (e) The distribution of time periods between neighbouring events, showing a strong correspondence to the distribution for a Poisson process with $\lambda = 1/10s$ (red).

We also investigate the distribution of events over time. One clear indicator of a Poisson distribution is an exponential probability density function in the time between neighbouring events. From the same 100 datasets as above, we find 326 time periods between adjacent events in the same dataset. Fig. S2e shows the normalised distribution of these times, along with the probability density function for a Poisson process with $\lambda = 1/10s$. The strong correspondence indicates that events occur independantly, as one might expect for events produced by radioactive decays or isolated cosmic ray impact events. We also note that the event rate found is sensitive to the lower cutoff for size of events located, as less sensitive methods will locate fewer events.

## IV. FURTHER EVENTS WITH HIGH TIME RESOLUTION

In order to study the evolution of events in greater detail, we performed RReCS experiments with a 3 $\mu$s interval between datapoints, and lasting 3 s total. Along with the event shown in Fig. 3 of the main text (Event A), two further events at this time resolution are shown in Fig. S3 (Events B and C).

All three events display the same overall timescales, showing an initial jump in a few tens of $\mu$s, a saturation to peak value in around 1 ms and an exponential decay back to baseline with a timescale of $\sim 25$ ms. However, each event shows a different location for the initial area of high error rates; Event A features an initial impact centered on the right side of the qubit patch, Event B features an initial impact on the bottom left side of the qubit patch, and Event C sees an initial impact on the top left of the patch. All three immediate impacts primarily affect a small patch of qubits, producing a jump from $\sim 4$ simultaneous errors at baseline to $\sim 10$ errors. Both events then spread through the chip, rising to a peak value around 1.5 ms after the initial impact. Finally, both events show the same exponential decay to baseline.

This provides further evidence that the initial localization is caused by a single point of impact and in a location that is independent between different events.

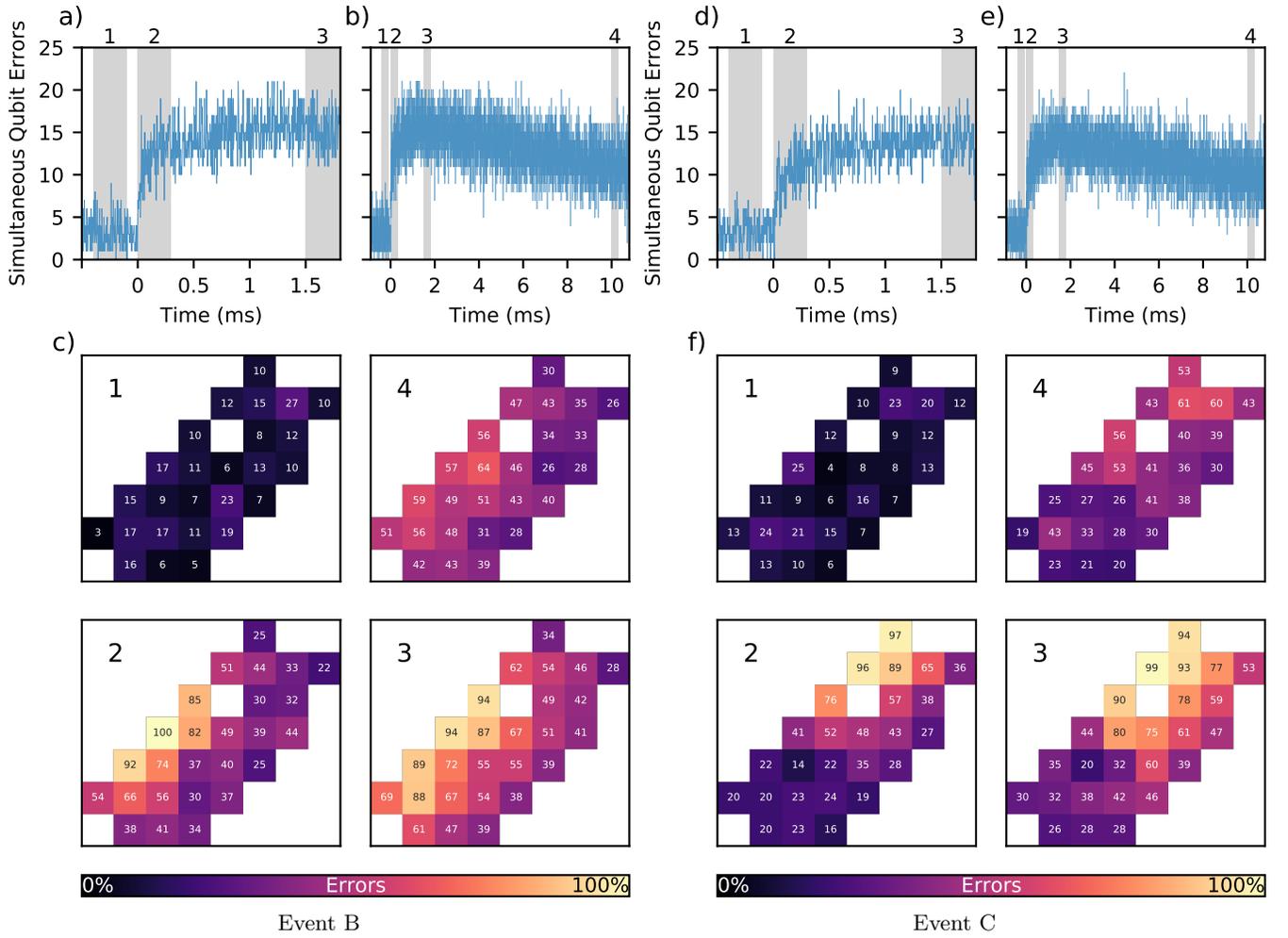

Figure S3. **Further highly time-resolved events** (a,b,d,e) Time slices from two highly time-resolved RReCS experiments with an interval of 3 $\mu$s between datapoints. Both events show the same timescales over their evolution despite differing in height. (c,f) Heat maps of the qubit patch, averaged over $300\mu s$ slices located (1) before the event, (2) at the initial impact, (3) after the rise to the peak value, (4) during the exponential tail of the event. Both events show initial high-error hot-spots in different locations, but later display elevated error levels throughout the chip.